\documentclass[reprint,amsmath,amssymb,aps,pra]{revtex4-1}
\usepackage[utf8]{inputenc}
\usepackage{graphicx}
\usepackage{dcolumn}
\usepackage{mathrsfs,amsmath}
\usepackage{bm}
\def\ph2{{\it p}-H$_2$}
\def\Am3{\AA$^{-3}$}

\begin{document}
\title{Second layer crystalline phase of helium films on graphite}
\author{Saverio Moroni}
\affiliation{CNR-IOM Democritos, Istituto Officina dei Materiali, and Scuola Internazionale Superiore di Studi Avanzati,
Via Bonomea 265, I-34136 Trieste, Italy}
\author{Massimo Boninsegni}
\email{m.boninsegni@ualberta.ca}
\affiliation{Department of Physics, University of Alberta, Edmonton, Alberta, T6G 2E1, Canada}
\date{\today}

\begin{abstract}
We investigate theoretically the existence at low temperature of a commensurate (4/7) crystalline phase of a layer of either He isotope on top of a $^4$He layer adsorbed on graphite. We make use of a recently developed, systematically improvable variational approach which allows us to treat both isotopes on an equal footing. We confirm that no commensurate crystalline second layer of $^4$He forms, in agreement with all recent calculations. Interestingly and more significantly, we find that even for $^3$He there is no evidence of such a phase, as the system freezes into an {\it incommensurate} crystal at a coverage lower than that (4/7) at which a commensurate one  has been predicted, and for which experimental claims have been made.   Implications on the interpretation of recent experiments with helium on graphite are discussed.
\end{abstract}
\maketitle
\section{Introduction}\label{intro}
The low temperature phase diagram of helium on graphite continues to intrigue both experimenters and theorists alike. Although the subject is now a few decades old \cite{bretz,hering,polanco,carneiro,ecke,lauter,freimuth,lauter2,greywall,greywall2}, and despite a considerable amount of investigation, some intriguing aspects have  not yet been fully elucidated, and remain highly debated. A chief example is the existence of a commensurate crystalline phase (henceforth referred to as 4/7) in the second layer of $^4$He, with a $\sqrt 7 \times \sqrt 7$ partial registry  with respect to the first
layer. Such a phase, occurring at coverages intermediate between the low-density superfluid and the high density incommensurate crystal, was first proposed by Greywall and Busch \cite{greywall,greywall2}, based on heat capacity measurements. Crowell and Reppy \cite{crowell,crowell2} in turn suggested that a ``supersolid'' phase \cite{note}, simultaneously displaying crystalline order and dissipation-less flow of $^4$He atoms, may exist at or near such a registered phase.
\\ \indent
To our knowledge, no direct, unambiguous experimental confirmation of the 4/7 phase of $^4$He on graphite has yet been provided. Furthermore, the most recent and reliable theoretical studies, namely first principle computer simulations based on state-of-the-art Quantum Monte Carlo (QMC) methods and realistic microscopic atom-atom and atom-surface potentials \cite{corboz,ahn}, have failed to confirm its existence, showing instead that the system remains in the superfluid phase at low temperature, at commensurate coverage. Nonetheless, the presence of such a phase still constitutes a working assumption in recent experimental studies of helium films adsorbed on graphite, where the contention of possible ``supersolid'' behavior, defined as coexistence of two different types of order in a single homogeneous phase, has been reiterated \cite{fukuyama,saunders}. 
\\ \indent
Assuming a two-dimensional (2D) first layer density between 0.118 and 0.122  \AA$^{-2}$ \cite{lauter2,fukuyama}, one ends up with a 2D density for the 4/7 upper layer close to 0.07 \AA$^{-2}$, i.e., very close to the estimated freezing density of $^4$He in two dimensions \cite{gordillo}. However, the fluid phase of an adsorbed layer can be stable at significantly higher density than in strictly two dimensions, as atomic motion in the transverse direction (mostly quantum mechanical in character at low temperature)  acts to soften effectively the repulsive core of the interatomic potential, ultimately responsible for solidification(see, for instance, Ref. \onlinecite{toigo}). Indeed, first principle simulations  yield evidence of second layer freezing  at a density $\sim 0.076$ \AA$^{-2}$, in the low temperature (i.e., $T\to 0$) limit \cite{corboz}. In any case, clearly caution should be exercised, as the physical proximity of all these putative phases means that the resolution of small energy differences is likely required, in order to map out the phase diagram correctly.
\\ \indent
No controversy exists as to whether the second layer is crystalline at 4/7 commensurate coverage, if it is made of atoms of the lighter ($^3$He) isotope (the first layer still of $^4$He atoms); indeed, in this case the experimental evidence is fairly robust (see, for instance, Ref. \cite{fukuyama2}). This is not entirely surprising, however, as $^3$He is well known to freeze at lower density than $^4$He, in spite of its lighter mass; in particular, theoretical studies \cite{motta} and experimental evidence \cite{bauerle} concur in assigning a 2D freezing density to $^3$He $\sim 0.06$ \AA$^{-2}$. In this case as well, one may expect an adsorbed layer to freeze at higher density, and interesting questions arise, namely {\em a}) if an intermediate, registered phase can intervene between the fluid and the incommensurate crystal, and {\em b}) how one can unambiguously identify  a commensurate phase, if it occurs inside a range of coverage in which an {incommensurate} one is thermodynamically stable.
\\ \indent
Clearly, a  cogent test of a reliable theoretical approach to the investigation of this system consists of reproducing the observed behavior of a second  layer of either helium isotope, offering useful insight as to why they might display different physics. The application of QMC techniques to Fermi system is, of course, hampered by the well-known sign problem; however, given the crucial role likely played by quantum statistics \cite{role}, it is necessary that its effect be included as accurately as possible.
\\ \indent
In this paper, we describe results of a theoretical study of the thermodynamic stability of a commensurate 4/7 crystalline phase of the second layer of helium on graphite at temperature $T$ = 0. We assume a first layer of $^4$He, whereas for the second layer we consider both helium isotopes. We used the accepted, standard model of a helium film adsorbed on graphite, based on realistic microscopic potentials to describe the interaction among helium atoms, as well as between the helium atoms and the graphite substrate.
\\ \indent
Our calculations make use of a recently developed variational approach \cite{ruggeri}, based on an iterative backflow renormalization, which has been shown to yield quantitatively {\em very} accurate ground state estimates for superfluid $^4$He (virtually exact in this case), and for $^3$He of quality at least comparable to that afforded by the most sophisticated fixed-node Diffusion Monte Carlo (DMC) calculations. 
The advantage of this methodology is that it allows us to treat both helium isotopes on an equal footing, as a variational calculation (which we carry out by means of standard Metropolis Monte Carlo) is not affected by a sign problem and therefore no {\em ad hoc} remedy is required to circumvent it (e.g., the well-known fixed node approximation), which inevitably degrades the relative accuracy of the fermion calculation with respect to the boson one. And, although a variational calculation is intrinsically approximate, the iterative scheme adopted here allows us {\em a}) to improve significantly over a standard trial wave function, in practice removing most of the variational bias, {\em b}) to gain important information on the physical effects that are missing in the initial {\em ansatz}. As a check of the physical predictions obtained using the variational approach we also carried out selected, targeted DMC calculations, which consistently confirmed the VMC results.
\\ \indent
 Our results show that no 4/7 commensurate crystalline phase of $^4$He exist, in agreement with previous calculations. Indeed, the ground state arising from the variational optimization shows no evidence of ordered atomic localization. On the contrary, $^3$He forms a triangular crystal, consistently with experimental observation; however, we find no evidence of ``pinning" of $^3$He atoms at specific adsorption sites, i.e., the crystalline ground state is found to be actually {\em incommensurate} with the underlying $^4$He layer. In other words, the physics of this layer is essentially that of the purely 2D system, i.e., it is not significantly affected by the underlying graphite substrate nor the $^4$He layer.
\\ \indent
The remainder of this manuscript is organized as follows: in sec. \ref{model} we describe the model Hamiltonian; in sec. \ref{meth} we offer a brief description of the methodology adopted in this work, and illustrate our results in sec. \ref{results}. 

\section{Model}\label{model}
The system is modeled as an ensemble of $N$ pointlike particles, $N_3$ of which are $^3$He atoms (half of either value of the spin projection), $N_4$ are $^4$He atoms. Both species obey the appropriate quantum statistics, namely Fermi (Bose) for $^3$He ($^4$He). When two layers of $^4$He are considered, $N_3=0$ and $N_4=N$, while for the case of a $^3$He layer on top of a $^4$He one, it is $N_3=4N_4/7$. The numerical results presented here are obtained with a number of particles $N$=132, the first layer consisting of  a triangular solid of 84 $^4$He atoms
with areal density $\rho_1=0.1195$ \AA$^{-2}$. Correspondingly, the density of the top layer is $\rho_2=0.0683$ \AA$^{-2}$.\\ \indent 
The system is enclosed in a simulation cell shaped as a  cuboid,  with periodic boundary conditions in all directions (but the length of the cell in the $z$ direction can be considered infinite for all practical purposes). The graphite substrate occupies the $z<0$ region.
\\ \indent 
The quantum-mechanical many-body Hamiltonian reads as follows:
\begin{eqnarray}\label{u}
\hat H = -\sum_{i\alpha}\lambda_{\alpha}\nabla^2_{i\alpha}+\sum_{i<j}v(r_{ij})+\sum_{i\alpha}U({\bf r}_{i\alpha}).
\end{eqnarray}
The first and third sums run over all particles of either species, with $\alpha=3,4$,  $\lambda_{3}\ (\lambda_{4})=8.0417 \ (6.0596)$ K\AA$^{2}$, and $U$ is the potential describing the interaction of a helium atom (of either species) with the graphite substrate, to which we come back below. The second sum runs over all pairs of particles, $r_{ij}\equiv |{\bf r}_i-{\bf r}_j|$ and $v(r)$ is the accepted Aziz pair potential \cite{aziz}, which describes the interaction between two helium atoms of either species. Such a potential ha been shown to afford a rather accurate description of the energetic and superfluid properties of $^4$He. 
\begin{figure}[h]
\includegraphics[width=\columnwidth]{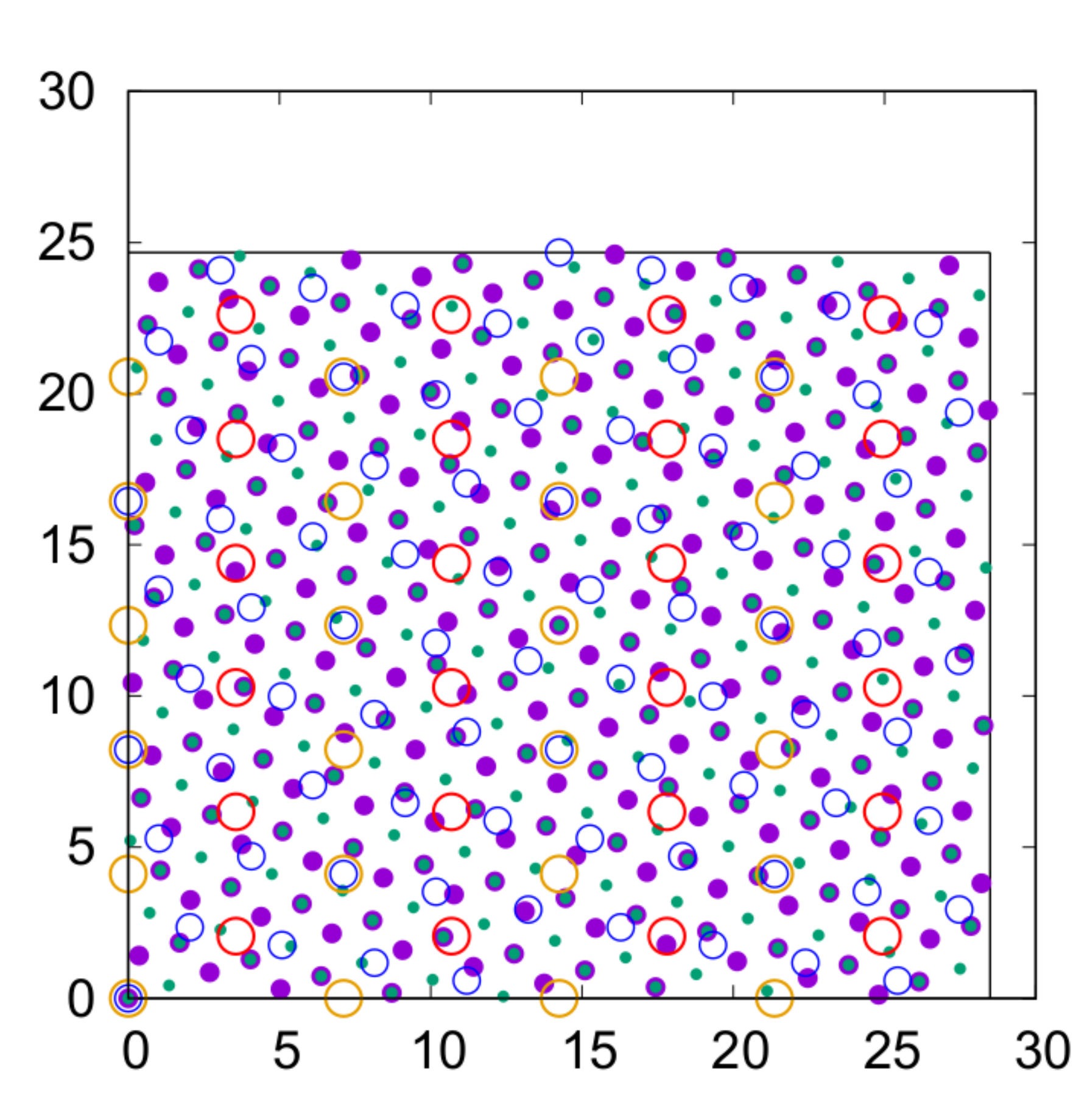}
\caption{The simulation cell. The filled circles are the A-- and the B--stacked
layers of graphite. The smaller open circles are the first layer lattice sites,
and the larger open circles are the second layer lattice sites (with
different colors for up-- and down--spins for $^3$He). The configuration
of lowest classical energy has the second layer is shifted by (1.26,0.45)~\AA.
}
\label{cell}
\end{figure}
\\ \indent
For the He-graphite interaction we consider two versions of the Carlos--Cole potential: 
the smooth, laterally averaged one \cite{cc}, or the corrugated anisotropic 6-12
potential \cite{cc2}.
The latter is tabulated for planar coordinates
within the $(x,y)$ unit cell of graphite as a function of the
distance from the surface, using 12 layers of carbon atoms within an 
in-plane cutoff of 55 \AA~ (43416 atoms). 
In the simulation the potential
is calculated by cubic interpolation of the tabulated values.
We ignore corrections 
for further C atoms. Even though they could be added perturbatively, they 
are very weakly dependent on $z$, and virtually identical in the liquid and 
crystalline phases.
As mentioned above, the system is periodic in $x$ and $y$, with simulation cell sides 
$L_x=28.4304489$~\AA~ and $L_y=24.6214914$~\AA. The cell (Figure~\ref{cell}) 
accommodates 268 sites of a slightly strained hexagonal 
lattice for graphite (the areal density of graphite is maintained at its
unstrained value corresponding to lattice parameters $a=2.461$~\AA~
and $c=6.708$~\AA). The anisotropic 6-12 potential includes this strain.
\\ \indent
Because we are also interested in the equation of state of an incommensurate crystalline top layer, we have also utilized in this study a simplified version of (\ref{u}), in which only the $N^\prime$ atoms in the top layer are explicitly included; they are assumed to move on a flat substrate, in the presence of a single--particle 1D potential ($v_{\rm eff}(z)$) which effectively accounts for both the Graphite substrate and the first $^4$He adlayer. We determine $v_{\rm eff}(z)$  as that whose ground--state
wave function is $\sqrt{\rho(z)}$, $\rho(z)$ being  the density profile  of He atoms in the second layer, computed using the full Hamiltonian (\ref{u}) with the corrugated potential. 
It has been shown \cite{ruggeri2} that the structural properties of $^3$He on
a smooth substrate, computed  with such an effective potential, are quantitatively very
similar to those on a corrugated substrate constituted by 
a solid layer of $^4$He, in turn adsorbed on (smooth
or corrugated) graphite. The advantage of this description, besides the computational speed-up arising from the reduction of the number of atoms that are explicitly modeled, is that on the effective smooth substrate
the density of a crystalline top layer can be varied continuously, in contrast to the case of an {\em explicit} solid $^4$He layer with fixed density, where the density is restricted
by the condition that the simulation cell accommodate both crystals.

\section{methodology}\label{meth}In this section we offer a description of the variational calculation, mostly focusing on the different wave functions utilized to describe the two phases of interest, namely crystalline and fluid. For a more thorough illustration of the approach, including technical details of its numerical implementation, we refer the reader to Ref. \onlinecite {ruggeri}.
We have utilized different trial wave functions for the 
system with liquid/solid $^4$He/$^3$He in the second layer, 
$\Psi_{L4}$, $\Psi_{S4}$, $\Psi_{L3}$ and $\Psi_{S3}$,
featuring a varying number of backflow iterations, until 
the result of interest (the stability of a given phase in our case) was deemedrobust against
further iteration.
All wave functions contain a common factor (optimized independently for each case)
\begin{eqnarray}\nonumber
&\Psi&_{0}(R)=\prod_{i<j}e^{-u_{\alpha\beta}(r_{ij})}
              \prod_ie^{-f_\alpha(z_i)}
              \prod_{i\in I}e^{-n_I(|{\bf r}^\perp_i-{\bf s}_i^{(I)}|)}\\
         &\times&
              \prod_{i\in I,j}e^{-m(|{\bf r}^\perp_i-{\bf h}_j|)}
              \prod_{i<j\in II}\prod_{k=0}^{n-1}e^{-w_k(q_{ij}^{(k)})}
\end{eqnarray}
where $\{{\bf r}_i\}=R$ are the coordinates of the He atoms; ${\bf r}^\perp_i$
are the $(x_i,y_i)$ components of ${\bf r}_i$; $\alpha$ and $\beta$ take the
value $I$ for the first layer and $II$ for the second layer;  
${\bf s}_i^{(\alpha)}$ are the in--plane components of the lattice sites of 
layer $\alpha$;
${\bf h}_i$ are the in--plane components of the centers of the hexagons
on the graphite surface; ${\bf q}^{(k)}_i$ are the coordinates of the $k$--th
iteration of backflow, given by ${\bf q}^{(k)}_i={\bf q}^{(k-1)}_i
+\sum_{j\neq i}\eta_k(q^{(k-1)}_{ij})({\bf q}^{(k-1)}_i-{\bf q}^{(k-1)}_j)$,
with ${\bf q}^{(-1)}_i={\bf r}_i$; the radial functions $u_{\alpha\beta}$, 
$f_\alpha$, $m$, $w_k$ and $\eta_k$ are suitable combinations of 
McMillan--like pseudopotentials and/or locally piecewise--quintic
Hermite interpolating functions \cite{natoli}, while the Nosanow factors
$e^{-n_\alpha}$ are gaussian functions; the function $m(r^\perp)$ is non--zero only 
for the corrugated graphite potential.
The wave function $\Psi_{L4}$ has an extra pair pseudopotential to
include a dependence on the $n$th iteration of backflow coordinates ${\bf q}_i^{(n)}$, i.e.,
\begin{equation}
\Psi_{L4}(R)=\Psi_0(R)\prod_{i<j\in II}e^{-w_n(q_{ij}^{(n)})}.
\end{equation}
On the other hand, 
$\Psi_{S4}$ has an extra Nosanow term in the in--plane components of ${\bf q}_i^{(2)}$ to describe a solid second layer, i.e.,
\begin{equation}
\Psi_{S4}(R)=\Psi_0(R)
\prod_{i\in II}e^{-n_{II}(|{\bf q}^{(n)\perp}_i-{\bf s}_i^{(II)}|)}.
\end{equation}
The wave function $\Psi_{L3}$ for the fluid phase of $^3$He has an extra Slater determinant of plane waves in the in--plane 
components of ${\bf q}_i^{(n)}$ (with twisted boundary conditions), namely
\begin{equation}
\Psi_{L3}(R)=\Psi_0(R)
\det_{ij} e^{i({\bf k}_j+\theta)\cdot{\bf q}^{(n)\perp}_i},
\end{equation}
whereas the crystalline wave function $\Psi_{S3}$ has an extra Slater determinant of Gaussian orbitals in the in--plane 
components of ${\bf q}_i^{(n)}$, centered at the lattice sites of the second layer, i.e., 
\begin{equation}
\Psi_{L3}(R)=\Psi_0(R)
\det_{ij} e^{-n_{II}(|{\bf q}^{(n)\perp}_i-{\bf s}_j^{(II)}|)}.
\end{equation}
Note that backflow coordinates are used only for atoms in the second layer.

\indent
The calculation of the ground state expectation values with the optimized wave function (corresponding to the $n$th backflow iteration) is carried out using a standard Metropolis Monte Carlo procedure, which of course does not suffer from any fermion ``sign" instability.
\\ \indent 
We now briefly discuss the correction of the energy estimates that we have implemented in order to account for the finite size of the simulated system.
We assume that the finite size effect on the kinetic energy is only
present for a fermion liquid (not for a bose liquid or a solid of
either statistics), due to the discreteness of the $k$--space shells
which enter the Slater determinant of plane waves. This is
eliminated (actually, strongly reduced) using twist--averaged boundary 
conditions \cite{lin}. The finite size effect
on the potential energy is estimated on a small subset of configurations
along the simulation
as the difference between the potential calculated with the minimum
image convention and the potential calculated with a large number of images.
The finite size correction turns out to be nearly identical for the liquid
and the solid phase of either He isotope.
\\ \indent
\section{results}\label{results}
\subsection{Stable phases at 4/7 coverage}
\begin{table}[h]
\begin{tabular}{|c|cc|}
\hline
 WF      &      VMC          &       DMC\\
\hline
                             & \multicolumn{2}{c|}{corrugated}       \\
$^4$He, $L$     & $-85.289\pm0.004$ & $-86.756\pm0.005$ \\
$^4$He, $S$     & $-85.244\pm0.004$ & $-86.688\pm0.006$ \\
$^3$He, $L$     & $-83.232\pm0.003$ & $-84.723\pm0.005$ \\
$^3$He, $S$     & $-83.323\pm0.003$ & $-84.769\pm0.005$ \\
                             & \multicolumn{2}{c|}{smooth}           \\
$^4$He, $L$     & $-85.037\pm0.002$ & $-85.684\pm0.004$ \\
$^4$He, $S$     & $-84.992\pm0.002$ & $-85.613\pm0.003$ \\
$^3$He, $L$         & $-82.980\pm0.002$ & $-83.630\pm0.003$ \\
$^3$He, $S$        & $-83.076\pm0.002$ & $-83.707\pm0.004$ \\
\hline
\end{tabular}
\caption{
Energy per He atom (in K) with either
the corrugated or the smooth He--graphite potential calculated 
in VMC and DMC using wave functions (WF) for liquid/solid $^4$He/$^3$He
in the second layer.
}
\label{table_1}
\end{table}
Table \ref{table_1} shows the
energy per He atom $E/N$, where $E$ is the total energy of the system,
yielded by the different variational wave functions for the two phases, namely liquid ($L$) and solid ($S$). The last column reports results obtained by means of DMC simulations, carried out by projecting out of the corresponding trial wave functions; the fixed-node approximation was used for those involving $^3$He. 
These results were obtained using the full Hamiltonian (\ref{u}), with either the corrugated anisotropic He-graphite potential or the laterally averaged, smooth one. 
\\ \indent
The first observation is that the quality of the wave function is
significantly better for the laterally averaged He-graphite potential, for which the difference between VMC and DMC results
is $\sim$ 0.65 K, than for the corrugated potential, where the
difference is $\sim$ 1.5 K, which amounts to roughly 1.8\% of the total energy. This may stem from the inadequacy of the 
part of the wave function describing correlations  between
first--layer atoms and graphite hexagons, which is expressed
through the two-dimensional, in-plane correlation function $m(r^\perp)$. Possibly, a more accurate  {\em ansatz} would be based on a fully three-dimensional function $m(r)$. 
\begin{table}[h]
\begin{tabular}{|c|cc|}
\hline
                                  & VMC              & DMC             \\
\hline
                                  & \multicolumn{2}{c|}{corrugated}     \\
$^4$He & -0.122$\pm$0.024 & -0.188$\pm$0.031\\
$^3$He &  0.251$\pm$0.017 &  0.128$\pm$0.026\\
                                  & \multicolumn{2}{c|}{smooth}         \\
$^4$He & -0.123$\pm$0.011 & -0.197$\pm$0.026\\
$^3$He &  0.264$\pm$0.011 &  0.148$\pm$0.019\\
                                  & \multicolumn{2}{c|}{effective}     \\
$^4$He & -0.154$\pm$0.002 & -0.179$\pm$0.003\\
$^3$He &  0.233$\pm$0.002 &  0.096$\pm$0.002\\
\hline
\end{tabular}
\caption{
Energy difference $\delta$ (see text) per second--layer atom (in K) 
between the liquid and solid phases of $^4$He/$^3$He 
in the second layer calculated in VMC and DMC using the full Hamiltonian (\ref{u}) with either
the corrugated or the smooth He--graphite potential, as well as with the effective potential $v_{\rm eff}(z)$ described in the text.
}
\label{table_3}
\end{table}
\\
\indent
On the other hand, the comparison between VMC and DMC estimates shows the same trend in both calculations; specifically, in no case is the prediction of relative strength of one phase with respect to the other made at the VMC level, reversed or even significantly quantitatively altered by DMC.
Indeed, as shown in Table \ref{table_3} the quantity $\delta\equiv (E_L-E_S)/N^\prime$, namely the energy {\em difference} per second layer atom between liquid and solid phases, is virtually unchanged (within statistical uncertainties) if either model of He-graphite interaction is used, for both isotopes and within either VMC or DMC. Moreover, $\delta$ is
consistently {\em negative} for $^4$He and {\em positive} for $^3$He. This remains true even if the calculation is based on the simplified version of model (\ref{u}) described above, making use of the effective potential $v_{\rm eff}(z)$.\\ \indent 
All of this
allows one to make a rather robust statement regarding the physical character of the ground state of the system in the case of an upper layer of either helium isotope. Specifically, the ground state of the second layer at coverage $\rho_2$ is a (translationally invariant) superfluid in the case of $^4$He, and a crystal for $^3$He.
As noted above, the value of the energy difference is essentially 
independent of the corrugation of the He--graphite potential, a fact that, while not particularly surprising for the case of $^4$He, for which the thermodynamic  equilibrium phase is a superfluid, is quite significant for the case of $^3$He, as it points to the equilibrium crystalline phase to be incommensurate, and thus scarcely affected by substrate corrugation. 
\\ \indent
Now, the fact that the energy per particle obtained using the wave function describing one phase (A) is lower than that for another phase (B), at a particular density $\rho$, is by itself no definitive proof that A is the true equilibrium phase at that density; for, it is in principle possible, in the case of a first-order phase transition, that $\rho$ fall within the region of coexistence of phases A and B. 
As we show below, this is not the case for the second layer density 
$\rho_2$, which corresponds to 4/7 commensurate coverage; indeed, $\rho_2$ does not fall within the liquid-crystal coexistence region, for a second layer of either helium isotope, as the calculations of the equation of state of the second layer show.
 
\subsection{Equation of state of the second layer}
We now discuss in detail the equation of state (EOS) for both a $^3$He and a $^4$He upper layer.
We compute the EOS by making use of the effective potential $v_{\rm eff}$ described above, representing both the graphite substrate and the first $^4$He adlayer. As explained above, the advantage of this approach is that 
the density of the crystalline top layer can be varied continuously, in contrast to the case of an {\em explicit} solid $^4$He layer with fixed density, where the density is restricted
by the condition that the simulation cell accommodate both crystals. We first present the results, and then discuss the expected accuracy of the approach.
\subsubsection{$^3$He upper layer}
\begin{figure}[h]
\includegraphics[width=\columnwidth]{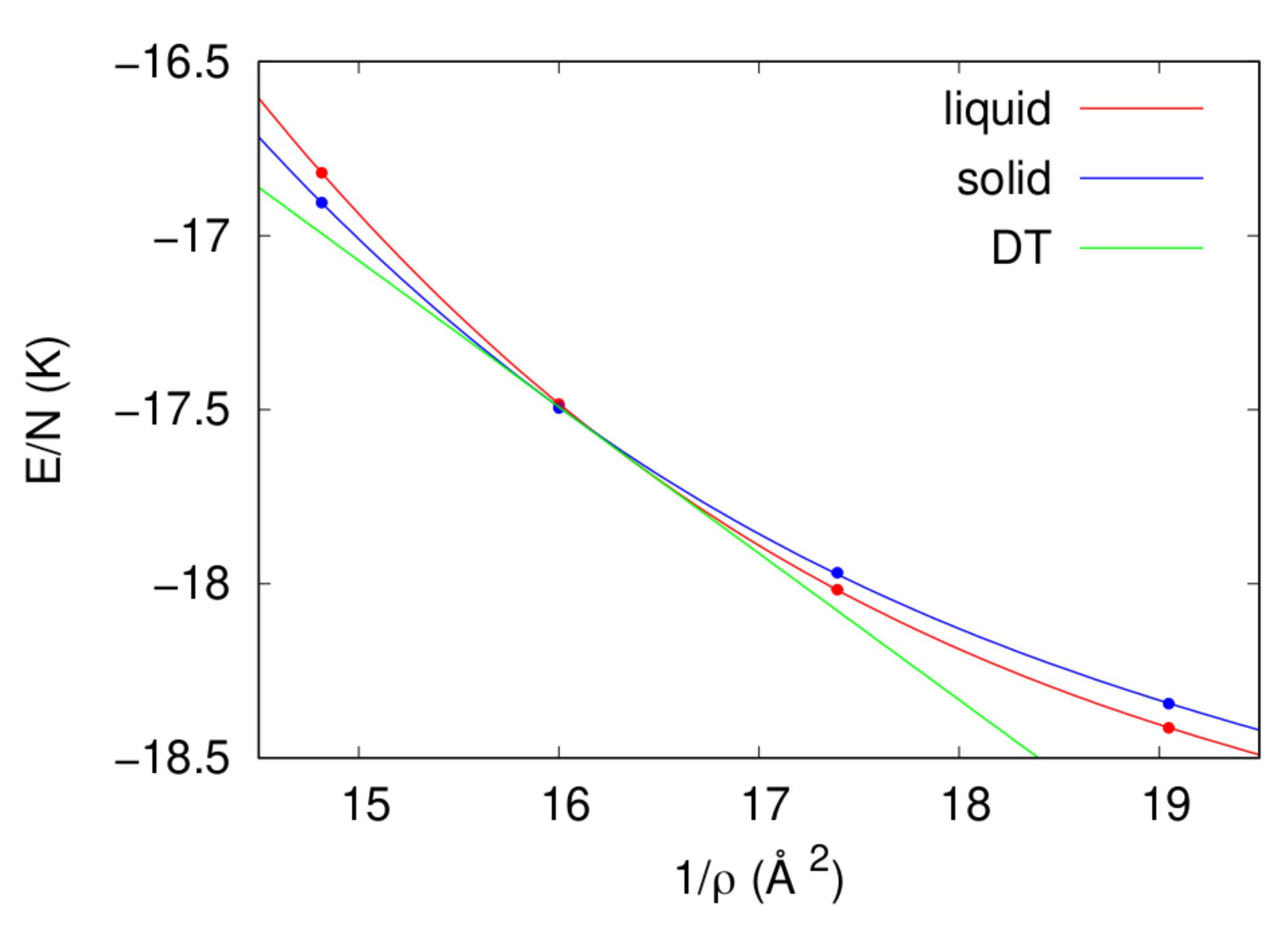}
\caption{{\em Color online}. EOS of a liquid and solid $^3$He layer adsorbed on a graphite substrate preplated with $^4$He, computed using the effective potential $v_{\rm eff}$ described in the text. The points are VMC energies
and the curves are cubic fits; the DT is also shown. 
The data pertain to simulations of 48 particles with periodic boundary conditions.
}
\label{fig_eos0}
\end{figure}

The EOS of a liquid and solid $^3$He upper layer, computed by VMC, are shown in Fig. \ref{fig_eos0}, together with the double 
tangent (DT) curve, $a+b/\rho$. 
\begin{figure}[h]
\includegraphics[width=\columnwidth]{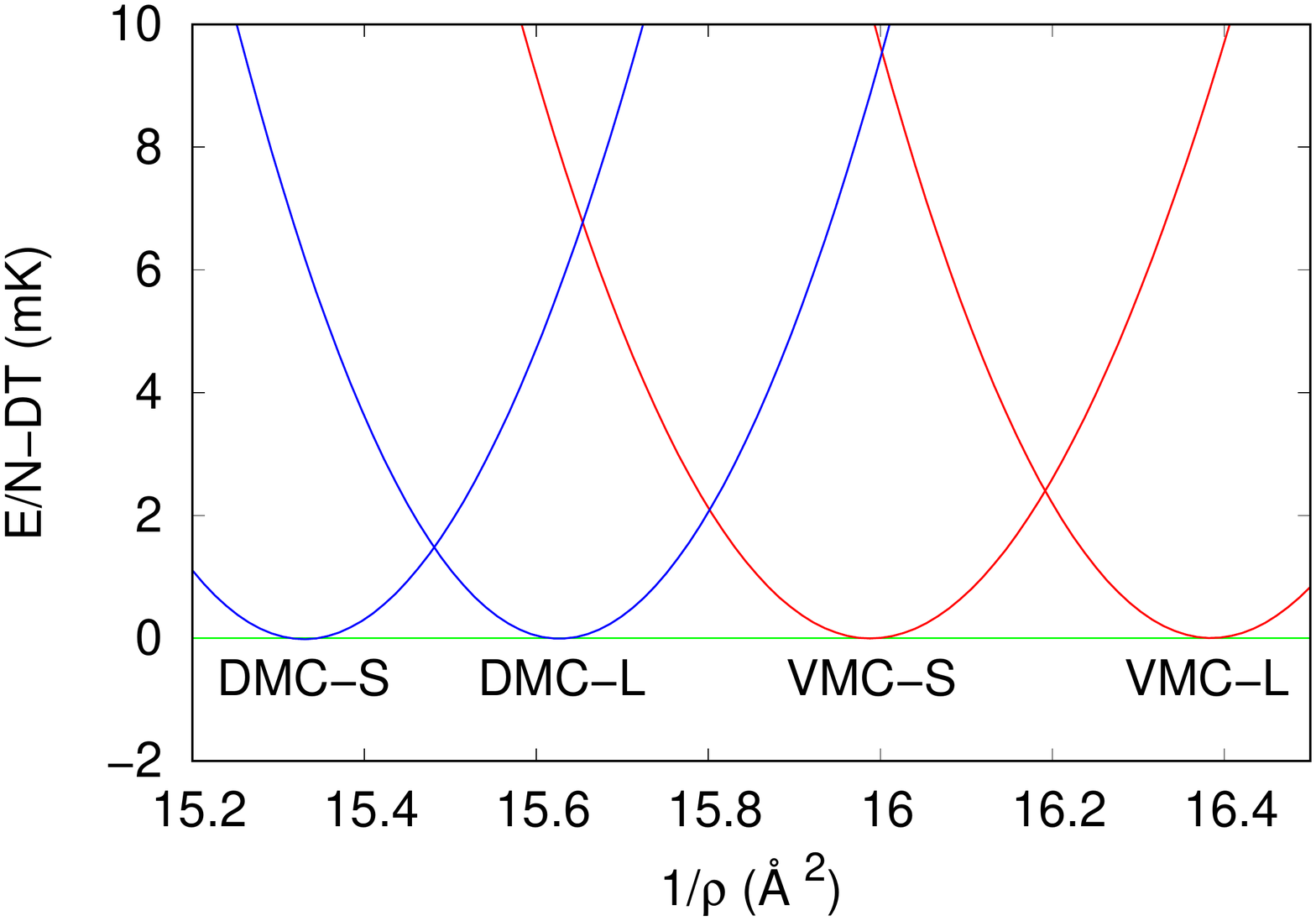}
\caption{{\em Color online}. Detail of Fig. \ref{fig_eos0} near coexistence,
with the DT subtracted, and the corresponding curves obtained with DMC energies.
}
\label{fig_eos1}
\end{figure}
The parameters $a$ and $b$ of the DT are determined by the condition 
that the difference with the DT vanish quadratically for both the liquid
and the solid EOS, as shown in Fig. \ref{fig_eos1}. The region of coexistence of fluid and crystal, computed by VMC, 
is given by the range of values of area per particle 15.99--16.38~\AA$^2$, or, equivalently,
0.061--0.062 ~\AA$^{-2}$ in density. In order to assess the quantitative accuracy of the VMC prediction, we performed fixed-node DMC simulations based on the optimized wave functions for both phases; as shown in Fig. \ref{fig_eos1}, the coexistence region is shifted to the area per particle interval 15.33--15.63 \AA$^2$, corresponding to the 0.064--0.065 \AA$^{-2}$ density range. Thus, our best estimate of the value of the melting density $\rho^\star$ is $\sim 0.065$ \AA$^{-2}$, still significantly lower than $\rho_2$, i.e., the density of the registered phase, equal to 0.0683 \AA$^{-2}$. Altogether, the agreement between VMC and DMC results is quantitatively excellent.
\\ \indent
\begin{figure}[h]
\includegraphics[width=\columnwidth]{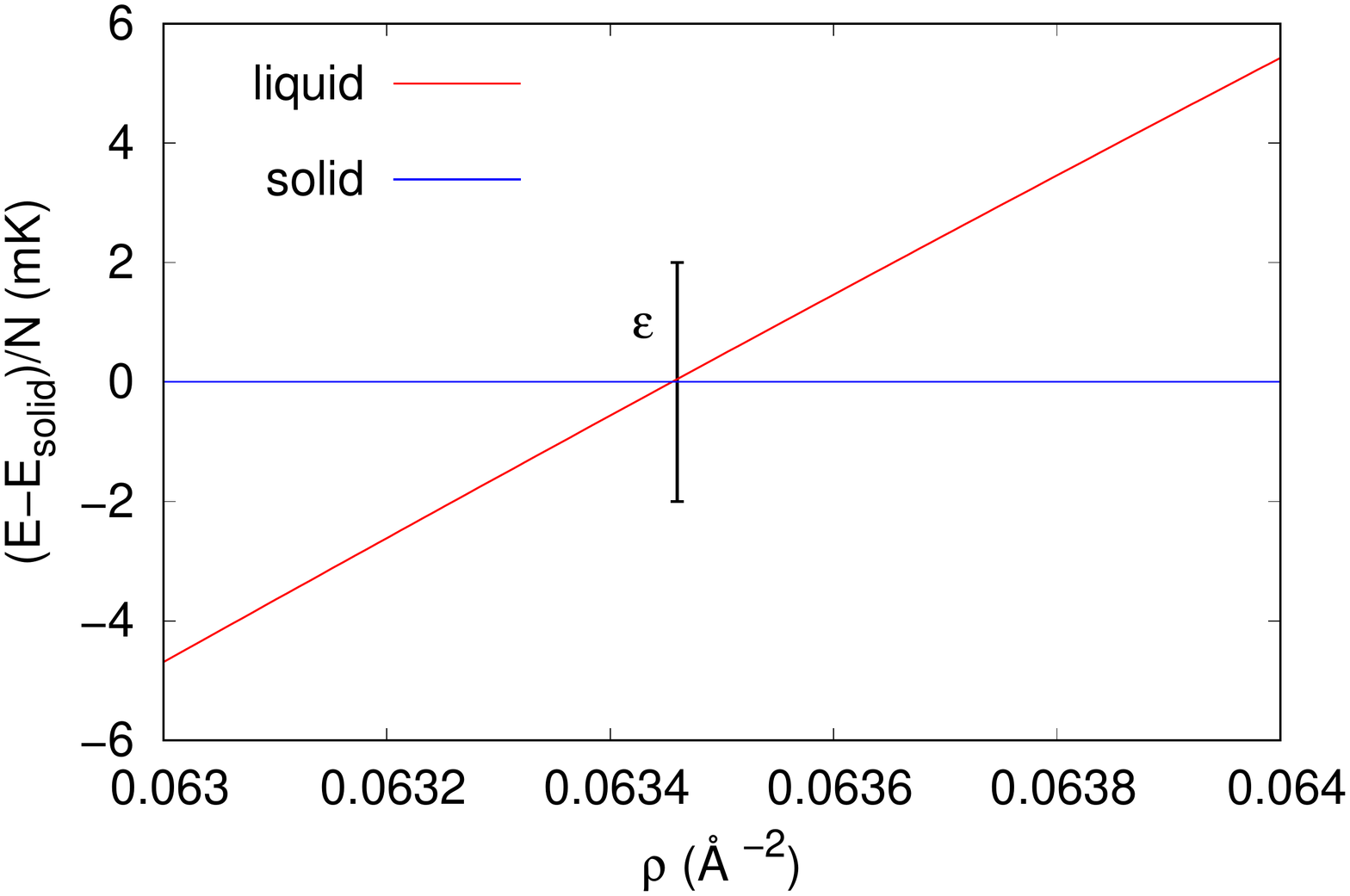}
\caption{Energy per particle for $^3$He on the smooth substrate relative
to that of the solid. The vertical bar shows the typical statistical error
of the data of Fig. \ref{fig_eos0}.
}
\label{fig_eos2}
\end{figure}
The uncertainty on the melting density $\rho^*$ can be estimated   through the energy difference
between liquid and solid (Fig. \ref{fig_eos2}), together with the typical size of the statistical
error of the data of Fig. \ref{fig_eos0}. 
The statistical uncertainty on the melting density is less than
0.001\AA$^{-2}$, which is significantly smaller than the difference between the density of
the 4/7 registered phase and $\rho^*$.
Figure \ref{fig_eos1} also shows that the liquid--solid energy difference 
spans a range $\lesssim$ 10 mK across the
coexistence region, much smaller than the liquid--solid differences
of $\sim$ 200 mK listed in Table \ref{table_3}. Therefore
the lower--energy phase at $\rho_2$ is definitely outside the coexistence 
region. 
\\ \indent
The results yielded by the model based on the effective potential suggest that freezing occurs to an incommensurate solid. 
Obviously,  we need to assess the extent to which the description based on $v_{\rm eff}$ is quantitatively representative of 
the model (\ref{u}), which explicitly includes the $^4$He atoms of the first layer. As mentioned above, the use of such an effective potential has been shown in previous work to be quantitatively reliable, and is not expected to alter significantly the predictions at which we have arrived using the effective potential.
Specifically, one should note that the liquid--solid
energy difference computed with the effective potential at density $\rho_2$ is slightly smaller in magnitude (by $\sim 35$ mK) than that computed with the explicit inclusion of the underlying $^4$He adlayer atoms, which has the effect of strengthening (albeit by a relatively small amount) the crystalline phase (no significant difference arises from the use of either the corrugated or the smooth helium-graphite interaction). Consequently, we may expect the melting density to be shifted to a slightly {\em lower} value if the full Hamiltonian (\ref{u}) is used, {\em a fortiori} validating our physical conclusion that the commensurate coverage $\rho_2$ falls well within the region of stability of the incommensurate crystal.
It is worth noting that our estimated freezing density is quantitatively consistent with
the highest density for which Bauerle {\em et al.} were able to measure
the spin susceptibility of a submonolayer liquid
$^3$He film adsorbed on a graphite substrate preplated by a monolayer of $^4$He \cite{bauerle}.
\\ \indent
We conclude by discussing the possibility that the crystalline phase of the $^3$He layer of density $\rho_2$ may be still registered with the underlying $^4$He layer, even though the density $\rho_2$ is inside the region in which the incommensurate crystal is energetically favored, at least according to our calculations based on the effective potential. This would be reflected by the ``pinning" of the $^3$He atoms at specific lattice locations, with a significant energy cost associated to, e.g., rigid relative translations or rotations of the upper layer with respect to the underlying one.
\\ \indent
In order to obtain a quantitative estimate of such pinning energy, we first considered two parallel, commensurate triangular lattices (first and second layers) spaced 3~\AA~ apart in the $z$-direction, and  computed the change in classical energy per He  atom associated to a rigid relative translation in the $(x,y)$-plane of one of the two lattices. The maximum energy change is in the range of few mK. We then carried out a DMC simulation of solid $^3$He over solid $^4$He, and found the change in the energy per particle between the highest-- and the lowest--energy classical 
configurations of the lattices to be reduced to few tens of mK, which is approximately ten times less than the
typical statistical uncertainty of this calculation. Such small values of the pinning energy do not, in our view, lend any  quantitative support to the
contention of an equilibrium crystalline phase of the upper $^3$He layer registered with the underlying $^4$He layer.
\subsubsection{$^4$He upper layer}
\begin{figure}[h]
\includegraphics[width=\columnwidth]{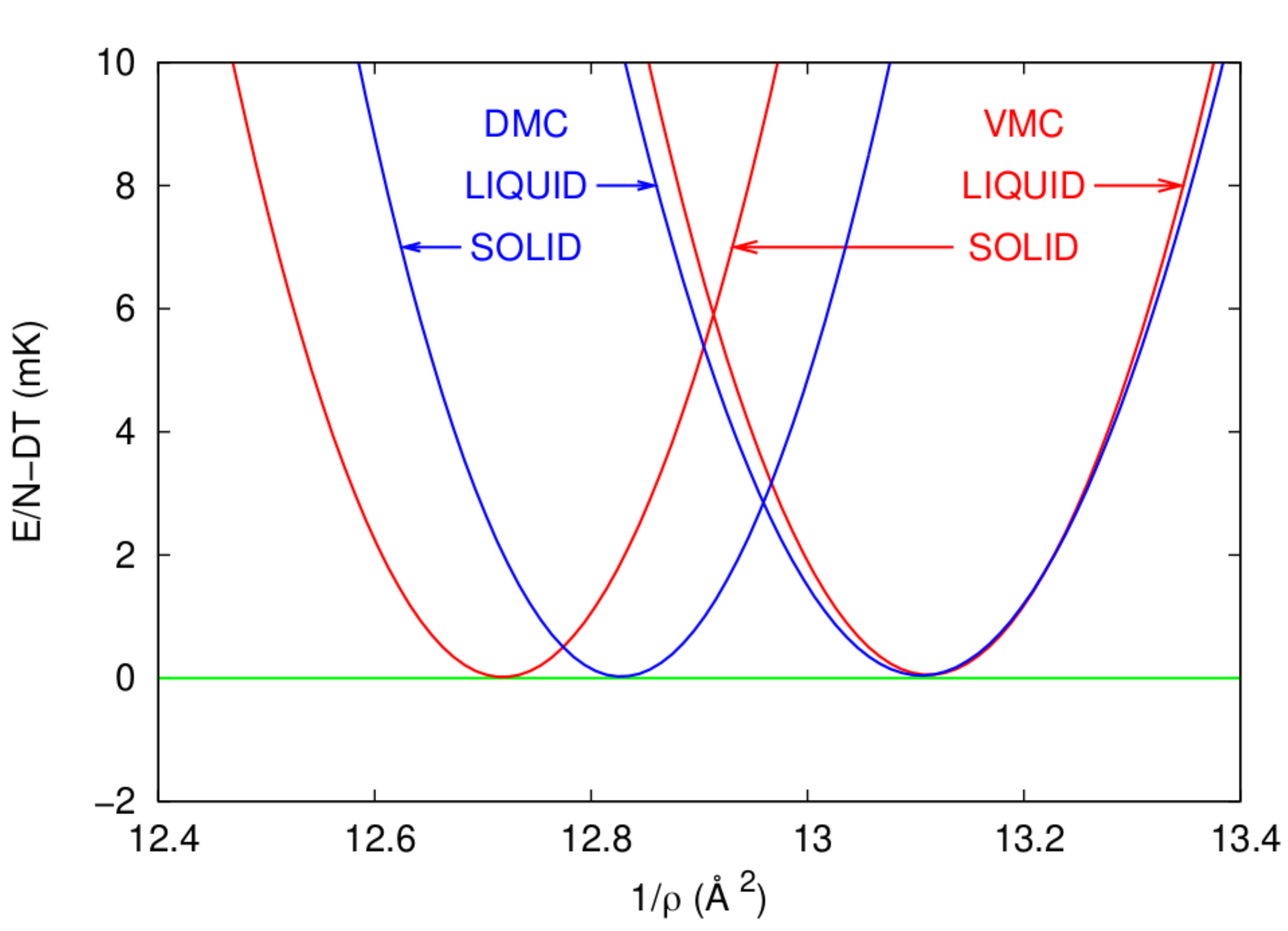}
\caption{{\em Color online}. Same as Fig. \ref{fig_eos1} but for a $^4$He upper layer. 
}
\label{fig_eos3}
\end{figure}
The same calculation has been carried out for a $^4$He second layer; figure \ref{fig_eos3} shows results  analogous to those of Fig. \ref{fig_eos1}. In this case, the coexistence region yielded by VMC is the density interval 0.076 -- 0.079 ~\AA$^{-2}$, which, as shown in Fig. \ref{fig_eos2}, is only slightly modified by subsequent DMC simulations, specifically shrinking to 0.076 -- 0.078 ~\AA$^{-2}$.
The freezing density is well {\em above} $\rho_2$, and is in excellent agreement with that estimate yielded by the finite temperature simulations of Ref.  \onlinecite{corboz}, explicitly including the $^4$He atoms of the first adlayer. This result gives us additional confidence in the use of the effective potential, as well as in the predictive power of the VMC methodology utilized here; it also supports the conclusion that no 4/7 crystalline phase exists, in agreement with the near totality of all numerical studies.
\section{Conclusions}\label{conc}
We have carried out a theoretical investigation of the possible existence of a 4/7 commensurate crystalline phase of the second layer of helium adsorbed on graphite. We considered both the case in which the upper layer comprises the same type of atoms as the first layer, namely $^4$He, as well as that in which the upper layer is formed by atoms of the lighter $^3$He isotope. We made use of a technique recently developed, aimed at studying the ground state of either Fermi or Bose systems by means of a variational (Monte Carlo) approach that affords high accuracy by iterative improvement of the wave function, and allows one to treat both isotopes of an equal footing.
\\ \indent 
The results obtained in this work constitute an additional piece of theoretical evidence against the existence of a commensurate crystalline phase in the second layer of $^4$He adsorbed on graphite. This is in agreement with the findings of essentially all the most recent theoretical calculations, based on first principle numerical simulations. It is worth restating that no {\em direct} experimental evidence of any registered crystalline phase of the second layer of $^4$He exists; rather, its presence has been proposed as a way to account for observed specific heat anomalies, for which, however, a different interpretation might have to be sought. Alternatively, the accepted microscopic theoretical model of $^4$He on graphite, which successfully accounts for most of the phenomenology, may have to be considerably revised (in ways that are not clear to us), should new and conclusive experimental evidence of a commensurate (4/7 or otherwise) phase arise.
 It has been suggested, however, that a 4/7 commensurate phase may also occur as a result of the first $^4$He layer forming a commensurate, rather than incommensurate crystalline phase as is commonly assumed \cite{ahn}.
\\ \indent
Our study also shows that no  commensurate phase exists if the second layer is formed by atoms of the lighter $^3$He isotope, a fermion, which undergoes crystallization into an incommensurate phase at coverages significantly lower than that of the putative 4/7 phase. The more general conclusion of this work is that the physics of the second layer of helium on graphite, of either isotope, is largely  independent of both the underlying $^4$He layer as well as of the graphite substrate; rather, it provides a close realization of the physics of $^3$He and $^4$He in two dimensions.
\section*{Acknowledgments}
This work was supported in part by the Natural Sciences and Engineering Research Council of Canada (NSERC). MB wishes to acknowledge the hospitality of the International Centre for Theoretical Physics in Trieste, Italy, where this  research was carried out.

\end{document}